# Semi-Supervised Multi-Organ Segmentation through Quality Assurance Supervision

Ho Hin Lee[a], Yucheng Tang**[a], Olivia Tang[a], Yuchen Xu[a], Yunqiang Chen[b], Dashan Gao[b], Shizhong Han[b], Riqiang Gao[a], Michael R. Savona[b], Richard G. Abramson[c], Yuankai Huo[a], Bennett A. Landman[a,d]

[a]Department of Electrical Engineering and Computer Science, Vanderbilt University, Nashville, TN, USA 37212;
[b]12 Sigma Technologies, San Diego, CA, USA 92130;
[c]Hematology and Oncology, Vanderbilt University Medical Center, Nashville, TN, USA 37235
[d]Radiology, Vanderbilt University Medical Center, Nashville, TN, USA 37235

(**Corresponding author: yucheng.tang@vanderbilt.edu)

**ABSTRACT**

Human in-the-loop quality assurance (QA) is typically performed after medical image segmentation to ensure that the systems are performing as intended, as well as identifying and excluding outliers. By performing QA on large-scale, previously unlabeled testing data, categorical QA scores (e.g. "successful" versus "unsuccessful") can be generated. Unfortunately, the precious use of resources for human in-the-loop QA scores are not typically reused in medical image machine learning, especially to train a deep neural network for image segmentation. Herein, we perform a pilot study to investigate if the QA labels can be used as supplementary supervision to augment the training process in a semi-supervised fashion. In this paper, we propose a semi-supervised multi-organ segmentation deep neural network consisting of a traditional segmentation model generator and a QA involved discriminator. An existing 3-D abdominal segmentation network is employed, while the pre-trained ResNet-18 network is used as discriminator. A large-scale dataset of 2027 volumes are used to train the generator, whose 2-D montage images and segmentation mask with QA scores are used to train the discriminator. To generate the QA scores, the 2-D montage images were reviewed manually and coded 0 (success), 1 (errors consistent with published performance), and 2 (gross failure). Then, the ResNet-18 network was trained with 1623 montage images in equal distribution of all three code labels and achieved an accuracy 94% for classification predictions with 404 montage images withheld for the test cohort. To assess the performance of using the QA supervision, the discriminator was used as a loss function in a multi-organ segmentation pipeline. The inclusion of QA-loss function boosted performance on the unlabeled test dataset from 714 patients to 951 patients over the baseline model. Additionally, the number of failures decreased from 606 (29.90%) to 402 (19.83%). The contributions of the proposed method are three-fold: We show that (1) the QA scores can be used as a loss function to perform semi-supervised learning for unlabeled data, (2) the well trained discriminator is learnt by QA score rather than traditional "true/false", and (3) the performance of multi-organ segmentation on unlabeled datasets can be fine-tuned with more robust and higher accuracy than the original baseline method. The use of QA-inspired loss functions represents a promising area of future research and may permit tighter integration of supervised and semi-supervised learning.

## 1. INTRODUCTION

The quality of medical image processing (e.g., segmentation) is affected by imaging quality, which can be influenced by both hardware-related and human-related artifacts [1]. The effectiveness and accuracy of image processing algorithms play an essential role in the usefulness and quality of medical image processing outcomes. To assess the image processing performance on a previously unseen dataset, quality assurance (QA) is typically performed to ensure the accuracy of the results. QA is a rich area of study and broadly consists of two main directions: subjective assessment, which is judged by human interaction, and objective assessment, which is decided by mathematical algorithms [2]. In medical image segmentation, human involved QA is still the de facto standard process to decide if the segmentation results are acceptable for the intended purposes. For instance, CT images with segmentation masks can be labeled with scores 0, 1 and 2 in a subjective manner and provide general information about the overlay image's quality. This quality score can be used to provide supplementary information in diagnosing abdominal organ disease via automatic algorithms such as deep learning.

With the onset of deep learning, large datasets are needed for training deep learning models; extensive engineering efforts are put into labelling and excluded outliers from large populations of medical images manually, especially CT images with low contrast [3]. Manually annotating organs with human expertise is preferred to ensure that the organs are labeled in the correct locations before used for training with supervised and semi-supervised learning methods [3]. However, manually performing annotations on medical images is time-consuming. Also, predictive models are dependent on the quality of the labels, and therefore suffer from outliers with low image quality. Image quality has become an important aspect to focus on and is shown to have a great impact on deep learning model prediction. In order to enhance the accuracy and efficiency of medical images analysis, previous research has been done for determining the image quality necessary to extract valuable information in medical perspectives.

Previously, assessing image quality of cardiac magnetic resonance (MR) images was proposed to detect the missing slice of the 3D MR images in cardiac scans and extract meaningful biomarkers from the missing slice with the use of convolutional neural networks [4]. The training dataset was extended with mis-triggering artifacts, using different levels of corruption in image quality to enhance the effect of image artifacts in training for data augmentation. On the other hand, an automated deep learning system has been created which uses retinal image quality as a feature to provide accurate diagnosis of diabetic retinography [5]. Also, uncertain chest x-ray image quality was leveraged to perform disease classification and lesion detection with deep Bayesian neural networks [6]. Producing uncertain predictions on medical cases leads to significant clinical value, while some of the cases need to be evaluated with physical exam or surgery to confirm [5, 6]. For medical image segmentation, an image-specific fine-tuning algorithm was proposed to make the convolutional neural network model adaptive to specific testing images and increase the generalizability of previously unseen data with the use of image quality [7]. New directions offer the potential to assess image quality to enhance the efficiency and accuracy of evaluating medical images. In recent years, with the use of deep learning, networks have been proposed to extract deep features and perform diagnosis in a robust manner, such as 3D U-net [8] and ResNet [9]. These networks have shown consistent interest and a great variety of usage in medical image segmentation [10, 11].

In this paper, we propose a semi-supervised learning method with a human-guided discriminator to determine the generalized quality of CT image segmentation and fine-tune the segmentation accuracy with the score prediction from the discriminator. The original 80 labeled 3D abdominal CT images along with the 2027 QA-only datasets were used for training the multi-organ segmentation model, along with the prediction of segmentation labels in each training epoch overlaid with the corresponding input image. Each overlaid 3D volume in the training phase was converted into a montage image and used as input into the discriminator to predict the segmentation quality at each training epoch. The prediction score from the discriminator was included in the loss and backpropagated to increase the model performance from QA-only data. 80 labeled datasets and 2027 unlabeled datasets are used in training the multi-organ segmentation model. The 3D Dice loss function is used for determining the difference between the predicted segmentation and the labels.

## 2. METHODS

In this paper, we propose a semi-supervised learning method for multi-organ segmentation with a 3D U-net and create a segmentation quality discriminator with ResNet-18 [9], which is presented in Figure.1.

### 2.1 Preprocessing

3D CT abdomen volumes are used as the raw datasets, and the volume part near the middle kidney was chosen via body part regression algorithms. Each value from the body part regression results represented the approximate location of the slice in the whole volume of 3D images. Slices with value near +12 are located around the duodenum, while slices with value -12 are near the heart. The slices with value -6 to +6 are extracted as a 3D volume and normalized to a resolution of 2x2x6 mm with dimensions of 168x168x64.

### 2.2 Network

Two independent networks are presented in Figure 1 that form the combined solution. For the multi-organ segmentation, a 3D U-net is used to extract deep 3D features with the encoder in U-net. Another network, ResNet18, is used as the discriminator to predict the segmentation quality score from score-labeled montage images.

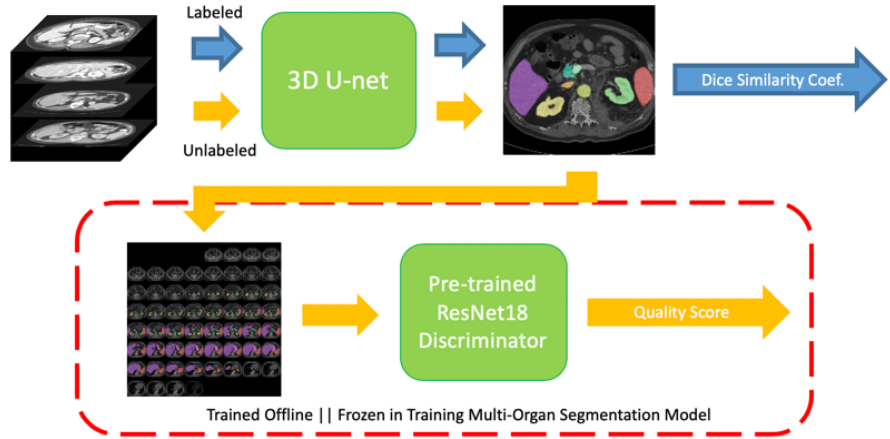

Figure 1. The full pipeline of semi-supervised learning-based segmentation is divided into two main processes: Multi-organ segmentation and quality discriminator module. All labelled and unlabeled data entered a 3D U-net to predict organ segmentation masks. Since Dice loss cannot be calculated from the unlabeled datasets due to the lack of ground truth, 2D slice alignment montage images are formed instead and predict segmentation quality score as the loss function. The quality scores backpropagated to the 3D U-net model to fine-tune the performance of segmentation on unlabeled datasets.

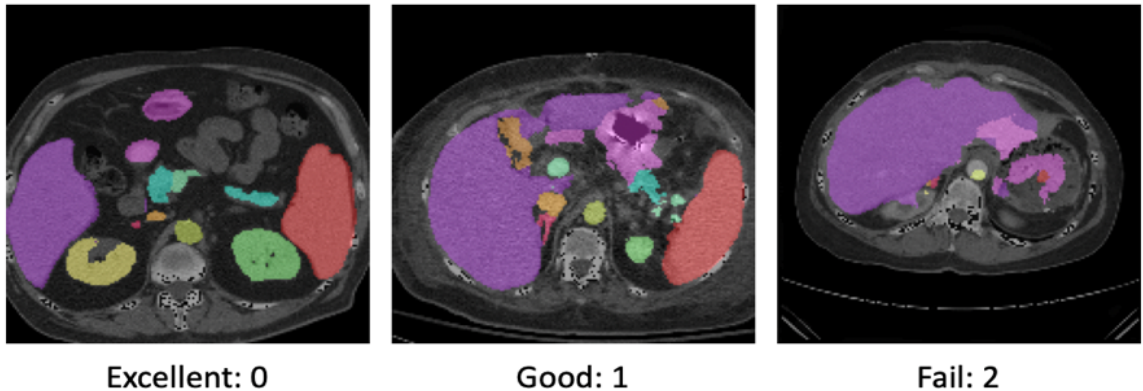

Figure 2: Respective images show the segmentation quality coded with 0 (success), 1 (errors consistent with published performance), and 2 (gross failure).

## 2.3  Segmentation Quality Discriminator

For unlabeled datasets, no ground truth masks were given for multi-channels to calculate the Dice similarity coefficients. After the segmentation mask prediction from 3D U-net, the segmentation mask was overlaid with its original input image and decomposed into 2D slices; these slices were aligned into a single montage image with the dimension 1344x1344. The discriminator was trained to determine the general quality of the segmentation predictions with the input of montage images. The input montages were downsampled to 256x256 for training to reduce the memory used. The discriminator obtained a high accuracy of 94% on 407 labeled test images with a ResNet18 trained with 1620 score-labeled montage images. A quality score was predicted and coded with 0 (success), 1 (errors consistent with published performance), and 2 (gross failure). After the model was trained, the discriminator model was frozen and put into the multi-organ segmentation pipeline to predict quality scores for unlabeled datasets. The score acted as the loss function for the multi-organ segmentation and was backpropagated from the frozen pre-trained network to fine-tune the performance on the datasets without labels.

## 2.4 Loss Functions for semi-supervised multi-organs segmentation

In the multi-organ segmentation pipeline, all labeled images and unlabeled images are input into the 3D U-Net to generate color segmentation prediction masks for different organs. For labeled images, 3D Dice loss was calculated for 12 organ channels, and all 12 anatomies were pushed through average Dice loss in the backpropagation and optimization process. We calculated the Dice loss with batch size 1 and all inliers and outliers were included to perform segmentation. The loss functions were:

*Multi-Sourced Dice Loss (MSDL):*

$$MSDL = -\frac{2}{A} \frac{\sum_{a=0}^{A} w \sum_{i=1}^{M} \sum_{j=1}^{N} A_{ij} P_{ij} + \in}{(\sum_{a=0}^{A} w \sum_{i=1}^{M} \sum_{j=1}^{N} R_{ij}^2 + \sum_{a=0}^{A} w \sum_{i=1}^{M} \sum_{j=1}^{N} P_{ij}^2 + \in)}$$

where $A$ is the number of anatomies, $w$ represents the variance between labels set properties and $P$ is the segmentation mapping for various organs. $\in$ ensures the stability of the loss function. Hence, $\in$ was used in computing the prediction and voxel value correlation. In the segmentation, 12 anatomies are adopted and the Dice loss function was iteratively optimized using Adam optimization.

*Mean Square Error Loss (MSEL):*

$$MSE = \frac{\sum_{i=1}^{N}(y_i - y_i^p)^2}{n}$$

For unlabeled data, the Dice loss function cannot be used due to the lack of ground truth segmentation masks. Therefore, the human-guided discriminator is used as a loss function and predicts a score for image segmentation quality. Montage images are created by slicing the 3D volume and aligning all 2D slices into one single color image with segmentations overlayed. MSE loss is used to calculate the difference between the human-labeled score and the predicted score.

## 3. DATA AND EXPERIMENTS

### 3.1 Data and platform

2107 total (both labeled and unlabeled) 3D abdomen CT images were used in the multi-organ segmentation pipeline. The data was retrieved in de-identified form from ImageVU under IRB approval. The volume of all datasets has dimensions of 168x168x64. 80 images from the datasets were manually labeled with all 12 anatomies. The remaining 2027 3D images were sliced and each of them was aligned as one single 2D montage image, which was used for training and validation in the discriminator module. Before inputting into the discriminator module, the montages are downsampled to 256x256 and overlaid with the mask prediction at each epoch. All 2027 montages were shuffled, and 80% of the datasets were used for training the discriminator, while another 20% of the montages were used for validation to assure the accuracy of the ResNet18 in comparing with score labels manually reviewed. For multi-organ segmentation, 80% of the original 3D volumes were randomly picked for training, and the remaining 20% were used for testing the performance of the segmentation model on labeled and unlabeled datasets.

### 3.2 Experiment Design
#### 3.2.1 Multi-Organ Segmentation

The multi-organ segmentation is performed as a baseline by changing the resolution from 1x1x3 mm to 2x2x6 mm and show the original state of art result for the 80 labeled and the 2027 unlabeled 3D datasets. 12 anatomies are segmented including spleen, kidney, gallbladder, esophagus, liver, stomach and pancreas [12]. After performing the segmentations, their montages are manually reviewed by experts, who determine the segmentation quality score (0,1,2) for unlabeled datasets. For the labeled datasets, Dice similarity coefficient were calculated to assess the performance of the segmentation.

### 3.2.2 3D U-net
For the multi-organ segmentation, a 3D U-net is used for extracting the deep features and increase the model capacity to reconstruct the image for mask prediction. The learning rate was set at 0.0001 and the network parameters are presented with the original paper published [8]. Multi or sourced Dice loss (MSDL) are used because of the multiple anatomies and segmentation mapping are the final output prediction from the network [13].

### 3.2.3 Discriminator module
After performing prediction of the segmentation mask, organ segmentation masks were overlaid with the original 3D volume and sliced into 2D. Each slice was aligned in a single montage image and down-sampled to 256x256, in order to input into the discriminator for determining the segmentation quality. The quality score was backpropagated through the ResNet, and montage processing back to 3D U-net model to fine-tune the model performance. The discriminator module was separately trained with 80% of the 2027 score labeled montage images. The remaining 20% of 2027 volumes were randomly separated into equal portions for testing and validation. A ResNet with 18 layers is used and performed regression to output a score for the segmentation quality. The learning rate was 0.0001 and the network parameters are presented with the original paper published [9]. MSE Loss was used to see the difference between the score label and the prediction score. For the model, validation was performed and epoch 198 model had the least root mean square error value.

### 3.2.4 Visual Quality Assessment
The quality score for each montage image was determined for labeling under several conditions: (1) The accuracy of the segmentation of liver, (2) kidney and (3) spleen. If the segmentation of liver, kidney and spleen were well located, the montage image were coded with score 0. Score 1 was coded for inaccurate segmentations of liver, spleen and kidney, as long as the segmentation mask could roughly define the location of each organ. For score 2, bad segmentations were allocated for each organ as presented in Figure 2.

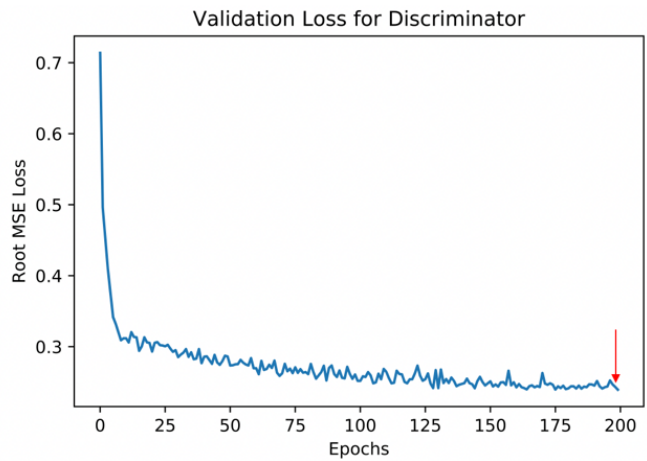

Figure 3: Validation loss as a function of training epoch and epochs 198 model is chosen with the least square root MSE error on the validation cohort

## 4. RESULTS
Testing loss was evaluated with the discriminator module as a function of epoch and was presented in Figure 3. The testing loss is essentially constant after 175 epochs, which is consistent with the validation performance. Based on the lowest validation loss value from validation curve, epoch 198 model was chosen with the red arrow shown in Figure 3. For the performance of multi-organ segmentation, a significant increase in segmentation quality for the unlabeled datasets is shown in Figure 4 and 951 patients compared to the baseline of 714 patients ($p<0.0001$). Additionally, the number of outliers has been reduced from 29.90% to 19.83% of all testing datasets. Quality of segmentations with each code are shown in Figure 5 and the accuracy for code 0 segmentation are increased comparing with the baseline. For the previously failed scans, the segmentation becomes more detailed, and the incorrect locations for the liver and spleen are corrected as shown in figure 5.

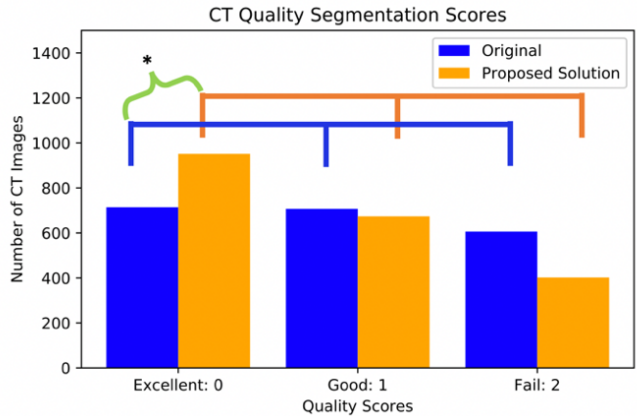

Figure 4: Significant increase in segmentation quality for unlabeled datasets is shown and reduce failure rate.
(*significant at p<0.0001)

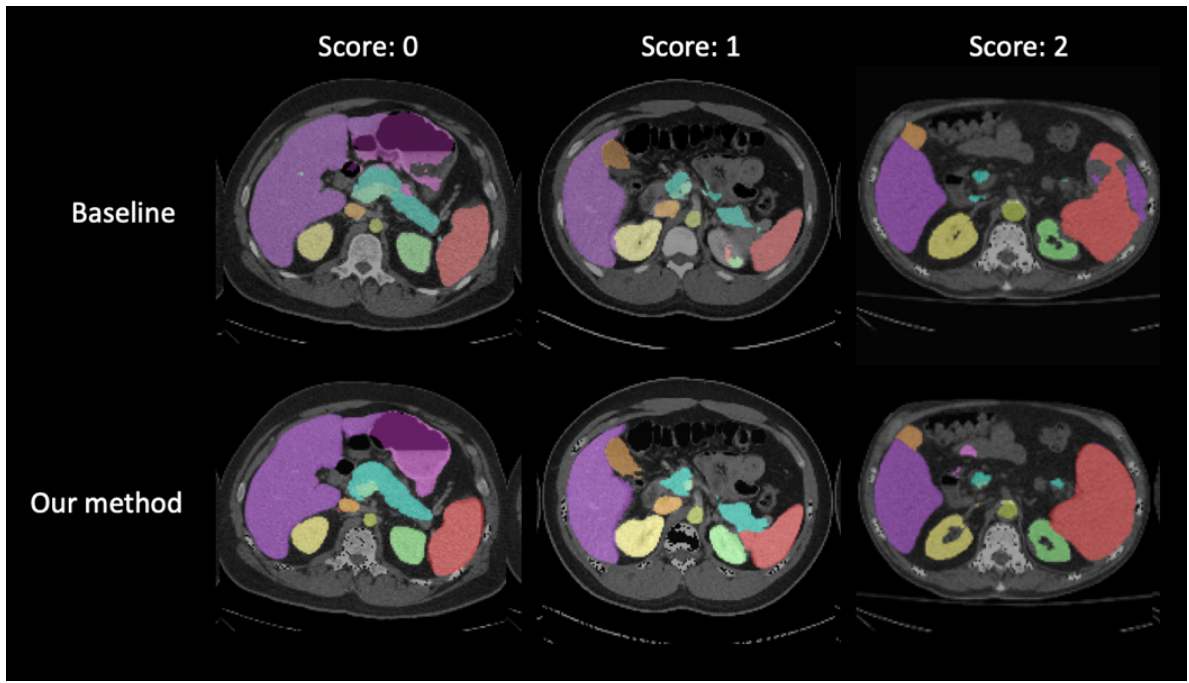

Figure 5: The segmentation mask overlay improved over the baseline method with our proposed method.

## 5. CONCLUSION

The proposed semi-supervised learning method with the integration of using QA scores to train the discriminator leads to more robust and effective segmentation performance. With the quality score predicted in each epoch for unlabeled data, the quality score acts as a loss function to provide feedback information for the original segmentation pipeline, improving the performance in outliers for segmentation. Hence, the use of QA-inspired loss functions represents an important perspective in supervised and semi-supervised learning, particularly for datasets with limited labels in medical image analysis.

## 6. ACKNOWLEDGEMENTS

This research is supported by Vanderbilt-12Sigma Research Grant, NSF CAREER 1452485, NIH 1R01EB017230 (Landman).This study was supported in part using the resources of the Advanced Computing Center for Research and Education (ACCRE) at Vanderbilt University, Nashville, TN. We gratefully acknowledge the support of NVIDIA Corporation with the donation of the Titan X Pascal GPU used for this research. The imaging dataset(s) used for the analysis described were obtained from ImageVU, a research resource supported by the VICTR CTSA award (ULTR000445 from NCATS/NIH) and Vanderbilt University Medical Center institutional funding. ImageVU pilot work was also funded by PCORI (contract CDRN-1306-04869).